\documentclass[a4paper, 12pt]{article}
\usepackage[english]{babel}
\usepackage[a4paper, inner=2cm, outer=2cm, top=3cm, bottom=3cm]{geometry}
\usepackage[dvipsnames]{xcolor}
\usepackage{mathtools}
\usepackage{pstricks}
\usepackage{caption}
\usepackage{graphicx}
\usepackage{amsmath}
\usepackage{amssymb}
\usepackage{mathcomp}
\usepackage{textcomp}
\usepackage{enumitem}
\usepackage{physics}
\usepackage{romannum}
\usepackage[doublespacing]{setspace}
\usepackage{subcaption}
\usepackage{hyperref}
\usepackage{xcolor}
\usepackage{booktabs}
\usepackage{cite}

\usepackage{academicons}

\newcommand{\orcid}[1]{\href{https://orcid.org/#1}{\textcolor[HTML]{A6CE39}{\aiOrcid}}}

\hypersetup{
  colorlinks   = true,
  urlcolor     = Blue, 
  linkcolor    = Black,
  citecolor    = Black
}

\newcommand{\thickhat}[1]{\mathbf{\hat{\text{$#1$}}}}

\definecolor{CrimsonGlory}{RGB}{179,0,30}

\title{On the Nonlocal Newtonian Cosmology}
\author{Raihaneh Moti \footnote{ r.moti@ipm.ir}\\
\textit{\footnotesize Department of Physics, Faculty of Science, Ferdowsi University of Mashhad, Mashhad, P.O. Box 1436, Iran} \\
\textit{\footnotesize School of Astronomy, Institute for Research in Fundamental Sciences (IPM), Tehran, P.O. Box 19395-5531, Iran}}
\date{}

\begin{document}
\maketitle
\pagenumbering{arabic}
\begin{abstract}
In the absence of an exact development of the cosmological models based on the Mashhoon nonlocal gravity, the Newtonian regime clarifies some aspects of it. To improve this model more reliable, going through some semi-Post-Newtonian considerations may be useful. One important feature to consider is the formation of horizons, which can be understood as a consequence of the finite interaction velocity. This highlights the importance of extending the integration across the entire space-time, rather than solely focusing on the spatial sector. In this context, we show that the density of effective dark matter would increase, while the density of Baryonic matter  decreases, during the deep matter-dominated era. This finding is in contradiction with the predictions of the standard model of cosmology and it raises concerns about the compatibility of using Tohline-- Kuhn kernel and considering the nonlocal effects as an effective dark matter in $\Lambda$CDM.
\end{abstract}

\section{Introduction}
The theory of general relativity is built upon the principles of equivalence and general covariance. The principle of equivalence stems from Galilean relativity, which asserts that the laws of motion remain unchanged in all inertial frames connected through Galilean transformations \cite{Blagojevic}. However, in order to attain general covariance, not only the laws of motion, but all the physical laws must be invariant. This is known as Einstein relativity and besides considering the light velocity as a finite standard quantity, it leads to the theory of special relativity (SR). Since the Galilean transformation is developed on the velocity addition rule, it is in contrast to SR. Consequently, the notion of inertial frames in SR has been redefined to encompass both the relative space-time and finite causal light speed.
In SR, the inertial frames are those connected by the Lorentz transformation instead of the Galilean transformation. Finally, the general covariance would be achieved by giving up the inertial constraint and writing the same physics in all frames. To do so, the Lorentz transformations are applied event by event to relate the instantaneous local inertial rest frame of the accelerated observer with the background global inertial frame\cite{Rahvar}. This is the \textit{locality hypothesis} and the local Lorentz invariance in the equivalence principle would not be obtained without it.

The validity of the locality hypothesis relies on the assumption that the physical processes described by the non-relativistic theory are point-like in Minkowski space-time, such as point particles and ray radiation. It becomes evident that any departure from the Eikonal ray approximation introduces an inherent scale and disrupts the notion of locality. If this scale is similar in magnitude to the length scale of the accelerated observer, the nonlocal effects cannot be disregarded\cite{Mashhoon1990a, Mashhoon1993}.

The Bohr--Rosenfeld point about the impossibility of instantaneous field measurement can be helpful to go beyond the locality hypothesis. They showed that the measurement of classical electromagnetic field at a given time $t$ by an ideal inertial observer involves an average over the past\cite{Bohr, Rovelli}. This idea has led to the development of nonlocal SR which is developed by Mashhoon\cite{Mashhoon1993, Mashhoon2011, Mashhoon Book}. In this theory, the accelerated observer measurements of the field contain an averaging process with a kernel that retains the history of its past acceleration. This noninertial effect at SR can also be extended to the gravitational interaction due to the equivalence of inertia and gravity.

Although, there are some suggestions to specify the kernel such as saving the causality and the universality of the model\cite{Mashhoon Book}, but there is no special one to fix its form. Without a precise development of nonlocal gravity (NLG), the Newtonian regime could shed light on certain aspects of it. In this extreme the nonlocal effects emerge as a modification of Poisson gravity equation,
\begin{equation}
\laplacian{\Phi(x)} + \int \chi(x-y) \laplacian{\Phi(y)} \dd[4]{y} = 4\pi G \rho(x) \label{MPEq}
\end{equation}
where $\rho(x)$ is the density of matter in Newtonian potential $\Phi(x)$ and the space-time coordinates are $x=(t,\vb*{x})$. The NLG kernel $\chi(x-y)$ should treat universally besides some other mathematical considerations. If the space of the model is restricted to the integrable ($L^1$) and square-integrable ($L^2$) kernels, rewriting the reciprocal second nonhomogeneous Volterra integral equation \eqref{MPEq} as
\begin{equation}
 4\pi G \rho(x)  + \int \mathcal{R}(x-y)  4\pi G \rho(y) \dd[4]{y} =\laplacian{\Phi(x)}
\end{equation}
where the kernel $\mathcal{R}(x-y)$ is the reciprocal of $\chi(x-y)$, facilitates the model more\cite{Mashhoon Book, Chicone2012}. Therefore, the modified Poisson equation is
\begin{equation}
\laplacian{\Phi} = 4\pi G (\rho_B+\rho_D)
\end{equation}
where convoluted nonlocal effects are
\begin{equation}
\rho_D (y) = \int \rho_B(x)\mathcal{R}(x-y) \dd[4]{x} \ . \label{DM density}
\end{equation}
These effects treat as a perturbation source to the classical matter. In the absence of the Baryonic matter ($\rho_B = 0$), the memory vanishes and the Newtonian mechanic is derived.

In the context of nonlocal electrodynamics, the presence of matter gives rise to the constitutive equation, and the kernel is derived from this equation which is based on the atomic physics of the underlying medium. However, in the case of gravity,  the kernel cannot be derived from the presence of matter but rather from the curvature of space-time, and there is currently no confirmed method to determine this kernel. In this situation  utilizing observational data appears to be a reasonable approach.

The phenomenological Tohline--Kuhn model recovers some observational aspects such as flat rotational curves of spiral galaxies\cite{Tohline1984, Kuhn1987, Bekenstein1988}. Therefore, the spherical symmetric Tohline--Kuhn kernel $q(r) = (4\pi\lambda_0 r^2)^{-1}$, where $r=\abs{\vb*{x}-\vb*{y}}$ could be an appropriate approximation at the Newtonian limit ($v/c \to 0$) where the action at distance is turned on. The $\lambda_0$ is the basic length scale of NLG which is related to the parameter of the Tohline--Kuhn model \cite{Mashhoon Book}. This is the main nonlocality parameter and the reciprocal kernel tends to zero as $\lambda_0$ tends to infinity \cite{Mashhoon2022}.

Therefore, the retarded effects are removed by\cite{Mashhoon Book}
\begin{equation}
\mathcal{R}(x-y) = \delta(x^0-y^0) q(\vb*{x}-\vb*{y}) \label{local kernel}
\end{equation}
and the gravitational memory becomes purely spatial
\begin{equation}
\rho_D (\vb*{y}) = \int \rho_B(\vb*{x})q(\vb*{x}-\vb*{y}) \dd[3]{x}  \ .  \label{DM density}
\end{equation}
 
The basic galactic scale $\lambda_0 \sim 1 \ kpc$  is the first parameter of the model. The other two, $a_0$ and $\mu_0^{-1}$ are the length scales parameters that moderate the short and long-distance behaviors of the $q(\vb*{x}-\vb*{y})$, respectively. Based on these considerations two kernels 
\begin{align}
& q_1(r) = \frac{1}{4\pi\lambda_0} \dfrac{1+\mu_0(a_0+r)}{r(a_0+r_0)} e^{-\mu_0 r} \ , \\
& q_2(r) = \frac{1}{4\pi\lambda_0} \dfrac{1+\mu_0(a_0+r)}{(a_0+r_0)^2} e^{-\mu_0 r}
\end{align}
are suggested\cite{Mashhoon Book}. The Tohline--Kuhn kernel would be recovered for $a_0 = \mu_0^{-1} = 0$ from these two. From Solar system data the lower bound $a_0\gtrsim 10^{15}$ was obtained\cite{Chicone2016}. Moreover, defining the dimensionless parameter $\alpha_0 = 2(\lambda_0\mu_0)^{-1}$ the rotational curves of the nearby galaxies and clusters of galaxies lead to\cite{Rahvar} 
\begin{equation}
\alpha_0 = 10.94 \pm 2.56 \quad, \quad \mu_0=0.059 \pm 0.028 \text{kpc}^{-1} \quad , \quad \lambda_0 \approx 3 \pm 2 \text{kpc} \ .
\end{equation}

Here, we explored the potential of using the Tohline-Kuhn kernel, which is derived from phenomenological considerations at the Solar system limits, to achieve consistent results at the cosmological limits as well. The findings indicate that for a simple cosmic model the nonlocal effects associated with this kernel do not yield valid results when compared to the standard cosmology model, $\Lambda$CDM.

We start by going one step beyond the Newtonian limit with restricting the space-time structure constant ($c$) to a finite value. This assumption saves the temporal integration in equation \eqref{DM density}. In Section \ref{second} we review the Nonlocal Newtonian Cosmology (NNC) with some considerations other than the standard form that is suggested in\cite{Mashhoon Book}. Then, the main model is introduced in section \ref{third}. Some concluding remarks are clarified in the last section \ref{fourth}.

\section{Nonlocal Newtonian Cosmology} \label{second}
The framework of Newtonian Cosmology relies on three fundamental equations: the continuity equation, the Euler equation, and the Poisson equation. In this model, the universe is likened to a vast gas cloud, which satisfies the hydrostatic equilibrium condition at the Newtonian limit. Thus, the fluid density of volume $V$ is $\rho$ and its internal energy is $U = \rho V $ \cite{Poisson, Motahareh}.
Since there is no heat flow in this homogeneous isotropic gas, the entropy exchange of this cloud which is
\begin{equation}
\dd{U}= -P \dd{V} + T\dd{S} \ ,
\end{equation}
vanishes. Therefore, the nonlocal effects emerge as a correction to the continuity equation, such that
\begin{equation}
\dot{\rho}_T(t) + 3 H(t) \left( \rho_T(t) + P_T(t) \right) = -\dot{\rho}_D(t) - 3 H(t) \left( \rho_D(t) +P_D(t) \right)
\end{equation}
where $\rho_T(t) \equiv \rho_B(t) + \rho_{\Lambda} $ and $P_T(t) \equiv P_B(t) + P_ {\Lambda}$ are the density and the pressure of the standard cosmology fluid with Baryonic contribution $(\rho_B(t),P_B(t))$ and cosmological constant contribution $(\rho_{\Lambda},P_{\Lambda})$. The dot denotes derivative with respect to the argument time $t$.

Assuming that the nonlocal effects are the memory of the Baryonic matter sector, $\rho_D(t) = \alpha(t) ~ \rho_B(t)$, indicates that the cosmological constant part does not have any contribution to the effective dark matter.

Same as the $\Lambda$CDM model, we assume that the components of the cosmic fluid is nothing but the radiation and matter. Thus, it can be shown that $P_D(t) = w_D \rho_D(t)$ where $w_D = \rho_r(t) / 3(\rho_r(t) + \rho_m(t) )$. And, as a result, the continuity equation is
\begin{small}
\begin{equation}
(1+\alpha(t)) \Bigl(\dot{\rho}_m(t) + 3H(t)\rho_m(t) + \dot{\rho}_r(t) + 4H(t)\rho_r(t) \Bigr) + \dot{\rho}_{\Lambda}(t)  + 3H(t) \rho_{\Lambda}(1+w_{\Lambda}) +  \dot{\alpha}(t)(\rho_m(t)+\rho_r(t)) = 0 \ .
\end{equation}
\end{small}
Here, we assume that the dust contributes the most in the matter sector. 

The Post-Newtonian model would be more reliable than Newtonian one to survey the behavior of the scale factor. In this model, the source of the Poisson equation is the \textit{active gravitational mass} $\rho + T^k_{\ k}$ not the fluid density alone \cite{Schutz}. Thus, we modified the NNC which is presented in\cite{Mashhoon Book} to the one with Poisson equation
\begin{equation}
\laplacian{\Phi} = 4\pi (\rho_T +3 P_T) 
\end{equation}
 besides the Euler equation
\begin{equation}
\pdv{\vb*{v}}{t} + (\vb*{v}.\nabla)\vb*{v} + \dfrac{\grad{P}}{\rho} + \grad{\Phi} = 0
\end{equation}
for a perfect fluid.

For a homogeneous isotropic expanding universe the space-time metric is
\begin{equation}
\dd{s}^2 =\dd{t}^2 - a(t)^2 \left( \frac{\dd{r}^2}{1-K r^2} + r^2 \right) \dd{\Omega}_2
\label{MF1}
\end{equation}
where $\dd{\Omega}_2 \equiv \dd{\theta}^2 +  \sin\theta^2 \dd{\phi}^2 $ and the scaling parameter of the Hubble law, $ a(t)$, satisfies the evolution equation
\begin{equation}
\dfrac{\ddot{a}(t)}{a(t)} = \dfrac{-4\pi G}{3} \bigl(2\rho_r(t) + \rho_m(t) - 2\rho_{\Lambda} + \alpha(t)(2\rho_r(t)+\rho_m(t)) \bigr)
\label{MF2}
\end{equation}
which is the second Friedman equation.

The first Friedman equation is the first integral of the second one,
\begin{equation} \label{1stFriedman}
\dfrac{\dot{a}(t)^2}{a(t)^2} = \dfrac{8\pi G}{3} \rho_B(t) (1+\alpha(t)) + \dfrac{8\pi G}{3} \rho_{\Lambda} - \dfrac{K}{a(t)^2} \ .
\end{equation}
The space curvature $K$ is the integration constant and it is assumed to be zero in the flat Newtonian limit.

The dynamical system is an apt tool to compare the time evolution of the system with the standard model of cosmology, $\Lambda$CDM. In the $\Lambda$CDM, the dimensionless densities $(\Omega_m,\Omega_r,\Omega_{\Lambda})$ are the parameters of the phase space and derived from three autonomous equations
\begin{align}
& \Omega'_m  = f_m(\Omega_m,\Omega_r,\Omega_{\Lambda}) \nonumber \\
& \Omega'_r = f_r(\Omega_m,\Omega_r,\Omega_{\Lambda}) \nonumber \\
& \Omega'_{\Lambda} = f_{\Lambda}(\Omega_m,\Omega_r,\Omega_{\Lambda})  
\end{align}
where the $f_i$s are functions of the $(\Omega_m,\Omega_r,\Omega_{\Lambda})$. They are raised from the Friedman equations besides the corresponding continuity equations of the matter, radiation, and cosmological constant.

For the present model, the autonomous condition is not satisfied by the simple choice
 $(\Omega_m,\Omega_r,\Omega_{\Lambda},\alpha\Omega_r,\alpha\Omega_m)$. Two points should be considered: proper choice of phase space parameters and proper splitting of the continuity equation. For the set $(\Omega_m,\Omega_r,\Omega_{\Lambda},\alpha\Omega_r,\alpha\Omega_m)$, customary splitting is not efficient. By  customary splitting, we mean the one which is used in the $\Lambda$CDM or the one based on the symmetry considerations\cite{splitting}. In the $\Lambda$CDM the splitting is based on the assumption that the fluid components do not interact with each other. The method suggested in \cite{splitting} is based on the invariance of the continuity equations under the transformation $\rho_i \leftrightarrow \rho_j$.

It can be shown that the set $(\thickhat{\Omega}_{m},\thickhat{\Omega}_{r},\Omega_{\Lambda})$ where the effective fluid density is
\begin{equation}
\thickhat{\Omega}_{i} = (1+\alpha(t)) \Omega_{i} \ ,
\label{parameter space}
\end{equation}
has a well-defined dynamical system. This choice is based on the physical assumption that the evolution of each component is only affected by its own past memory as an effective dark matter. Thus, if the first Friedman equation is rewritten as
\begin{equation}
\thickhat{\Omega}_{m} + \thickhat{\Omega}_{r} + \Omega_{\Lambda} = 1
\end{equation}
then, the set
\begin{align}
&\thickhat{\Omega}_{m}' = - \thickhat{\Omega}_{m} \Bigl( 1- (\thickhat{\Omega}_{m}+2\thickhat{\Omega}_{r} ) + 2 \Omega_{\Lambda} \Bigr) \ , \nonumber \\
& \thickhat{\Omega}_{r}' = - \thickhat{\Omega}_{r} \Bigl( 2- (\thickhat{\Omega}_{m}+2\thickhat{\Omega}_{r} ) + 2 \Omega_{\Lambda} \Bigr) \ ,  \\
& \Omega'_{\Lambda} =  -\thickhat{\Omega}_{m}' -\thickhat{\Omega}_{r}' \nonumber
\end{align}
would be autonomous, where the prime denotes derivative with respect to the $\ln{a}$.
The fixed points of this system are located at the zeros of each $f_i(\thickhat{\Omega}_{m},\thickhat{\Omega}_{r},\Omega_{\Lambda})$, where
\begin{align}
&f_m(\thickhat{\Omega}_{m},\thickhat{\Omega}_{r},\Omega_{\Lambda}) = - \thickhat{\Omega}_{m} \Bigl( 1- (\thickhat{\Omega}_{m}+2\thickhat{\Omega}_{r} ) + 2 \Omega_{\Lambda} \Bigr) \ , \nonumber \\
& f_r(\thickhat{\Omega}_{m},\thickhat{\Omega}_{r},\Omega_{\Lambda})  = - \thickhat{\Omega}_{r} \Bigl( 2- (\thickhat{\Omega}_{m}+2\thickhat{\Omega}_{r} ) + 2 \Omega_{\Lambda} \Bigr) \ , \\
& f_{\Lambda}(\thickhat{\Omega}_{m},\thickhat{\Omega}_{r},\Omega_{\Lambda})  =  -\thickhat{\Omega}_{m}' -\thickhat{\Omega}_{r}' \ . \nonumber 
\end{align}
Same as the standard model, the critical points are $\textbf{P}_m= (1,0,0)$, $\textbf{P}_r= (0,1,0)$ and $\textbf{P}_{\Lambda}= (0,0,1)$ where $\textbf{P}_i: (\thickhat{\Omega}_{m},\thickhat{\Omega}_{r},\Omega_{\Lambda}) $.

The eigenvalues of the stability matrix determine the stability at each point. At  $\textbf{P}_m$ the eigenvalues are $\lambda_m: (3,-1,0)$, at $\textbf{P}_r$ they are $\lambda_r: (4,1,0)$ and at $\textbf{P}_{\Lambda}$, $\lambda_{\Lambda}: (-4,-3,0)$. Clearly, the only stable fixed point is the $\textbf{P}_{\Lambda}$ and the others are unstable critical points.

\begin{figure}
    \centering
    \begin{subfigure}{\textwidth}
     \centering
		\includegraphics[width=0.8\textwidth]{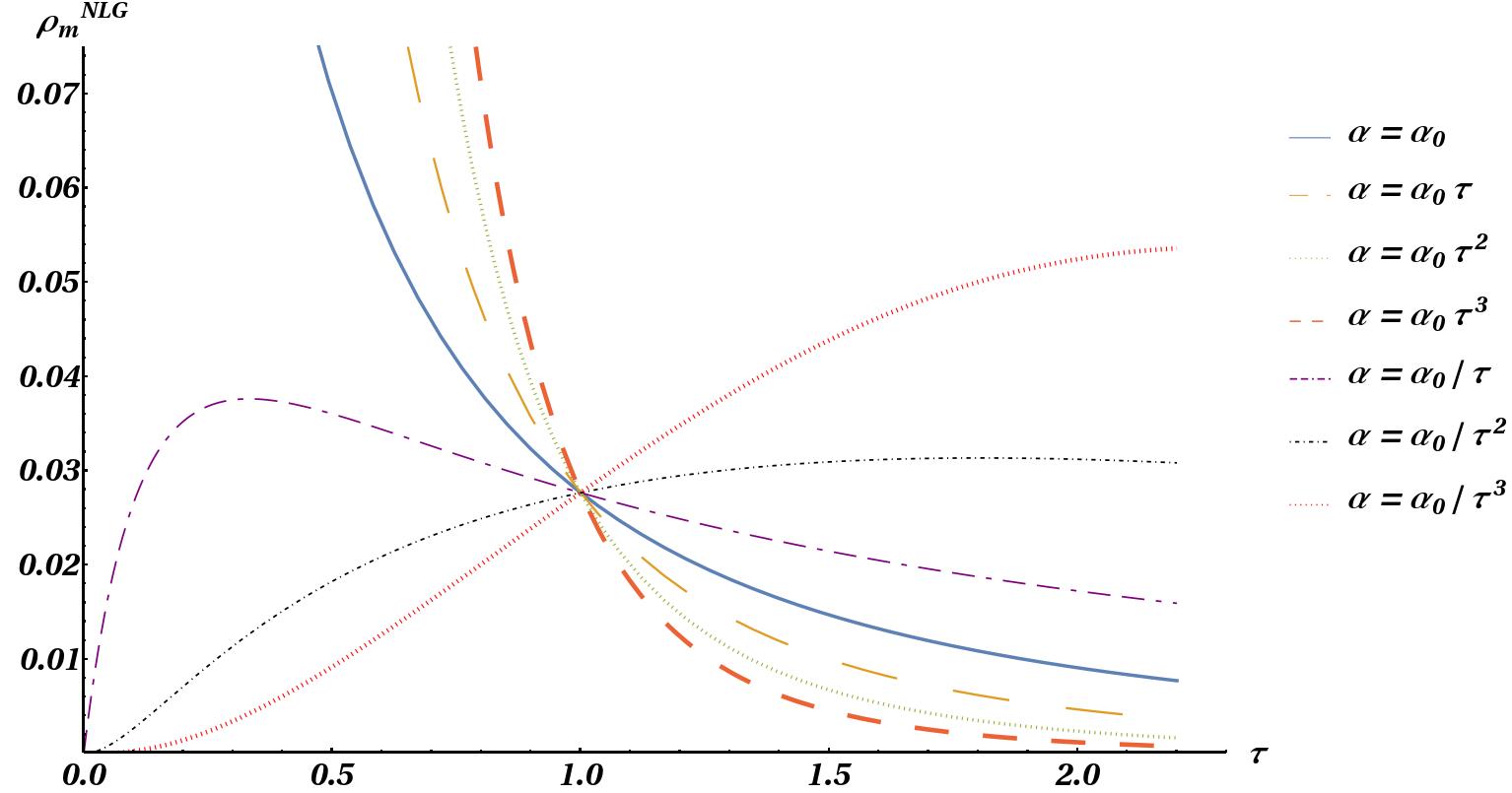}
	\end{subfigure}
	\\[1cm]
    \begin{subfigure}{\textwidth}
     \centering
		\includegraphics[width=0.8\textwidth]{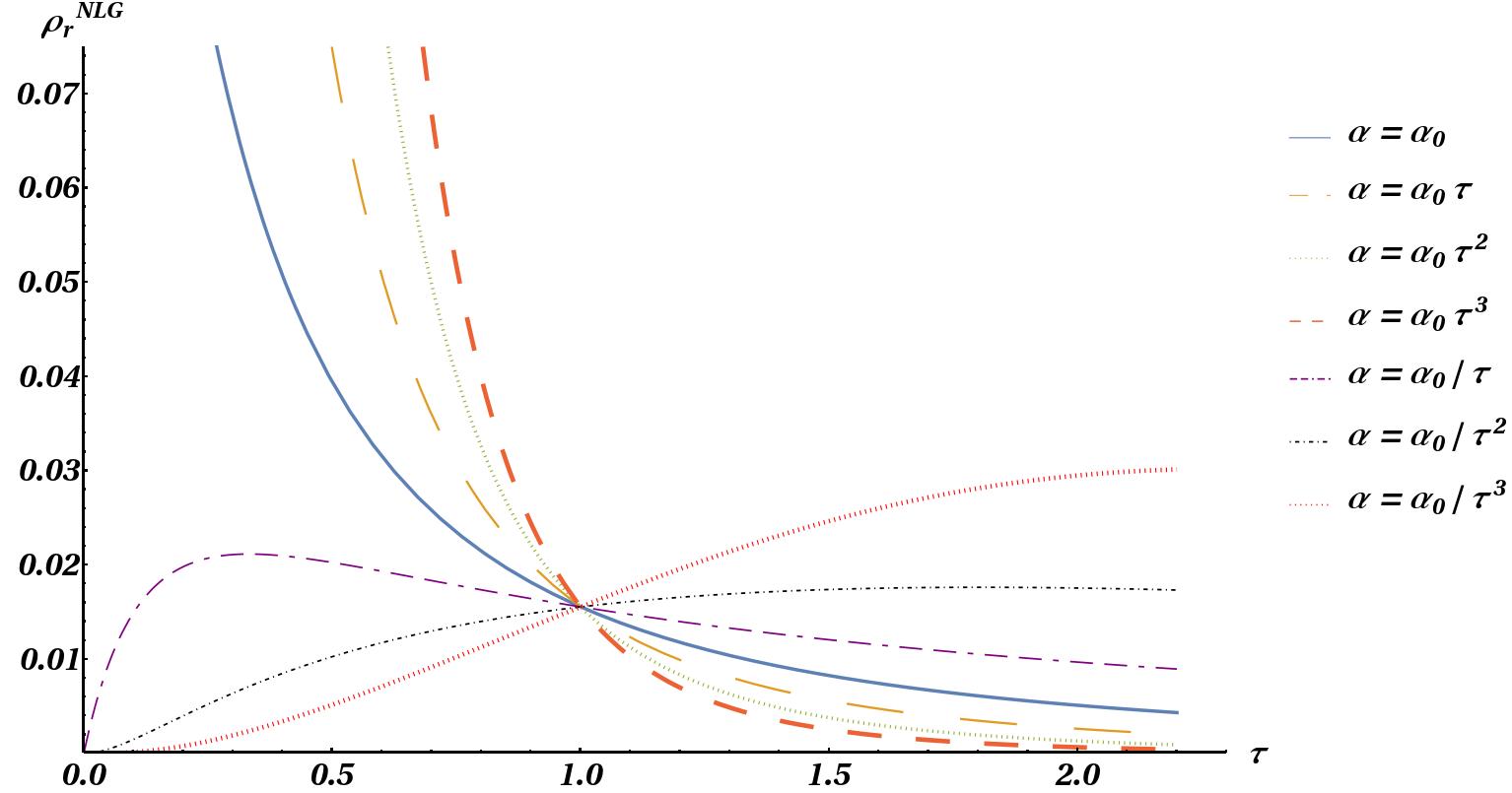}
	\end{subfigure}
	\caption{\small{The evolution of the matter density (top) and the radiation density (bottom) with respect to $\tau=t/t_0$.}}
	\label{density}
\end{figure}

It should be noted that the mentioned points are in the effective density space, not the density space. This could be clear more if we probe the solutions of the \eqref{1stFriedman} and
\begin{align}
& \dot{\rho}_m(t) + 3 H(t) \rho_m(t)  = -\dfrac{\dot{\alpha}(t)}{1+\alpha(t)}\rho_m(t)  \label{Ceq-matter}\\
& \dot{\rho}_r(t) + 4 H(t) \rho_r(t)  = -\dfrac{\dot{\alpha}(t)}{1+\alpha(t)}\rho_r(t) \label{Ceq-radiation}\\
& \dot{\rho}_{\Lambda} = 0 \ .
\end{align}
For a general nonlocality coefficient $\alpha(t)$ the density of the matter and the radiation sectors are
\begin{equation}
\rho_i(t) = \dfrac{a_0^{n_i} \rho_{0i}(t)}{a^{n_i}(t) (1+\alpha(t))}
\end{equation}
where $n_m =3$ and $n_r=4$. Although the effective density $\thickhat{\rho}_i = (1+\alpha(t)) \rho_i(t)$ treats same as its $\Lambda$CDM counterpart, the behavior of the  $\rho_i(t)$ depends on the chosen $\alpha(t)$, figure \ref{density}. 

Unlike the density fluid, the scale factor treats same as the $\Lambda$CDM at each era and at the $\textbf{P}_m$ and $\textbf{P}_r$ they are
\begin{align}
& a_m(t) = a_{0m} \Bigl( \dfrac{18 \pi G \rho_{0m}}{3} \Bigr)^{1/3} (t_0 \pm t)^{2/3}  \nonumber \\
& a_r(t) = a_{0r} \Bigl( \dfrac{32 \pi G \rho_{0r}}{3} \Bigr)^{1/2} ( t_0 \pm t)^{1/2} \quad , \quad a_{0r} \Bigl( \dfrac{32 \pi G \rho_{0r}}{3} \Bigr)^{1/2} ( -t_0 \pm t)^{1/2}
\end{align}
respectively.

\section{Beyond Nonlocal Newtonian Cosmology} \label{third}
To go beyond the Newtonian limit, let us first examine a quasi Post--Newtonian limit, by considering the invariance of the theory on the Lorentz transformation instead of the Galilean one. This is nothing but assuming that the constant of space-time structure ($c$) is the standard of the interaction velocity and it is finite.
Bounding this constant leads to horizon formation which restricts the causal region. Thus, the condition $\rho_D(t)=\alpha(t)\rho(t)$ which is satisfied at Newtonian limit could not describe this model anymore, since the nonlocal effects are from the whole of space-time history, not only the space sector.

The general nonlocal effects are defined as
\begin{equation}
\rho_D (y') = \int \rho_B(x')\mathcal{R}(x'-y') \dd[4]{x'} \ , \label{our-suggestion}
\end{equation}
where for a standard homogeneous cosmic fluid, the spatial dependence of the density is faded by the homogeneity condition, i.e $\rho(x')=\rho(t')$.
We assume that the temporal nonlocal effects are negligible to compare the results with NNC. Thus, the choice $\mathcal{R}(x'-y') = q_0(\qty|\vb*{x'-y'}|) \equiv q_0(r)$ is suitable for present model. It should be noted that unlike the kernel \eqref{local kernel}, here there is no action at distance anymore and as a result we do not consider the coefficient $\delta(x^0-y^0)$.

For $q_0(r) = \alpha_0 \mu_0 e^{-\mu_0 r} (1+\mu_0 r) / 8\pi r^ 2$, the effective dark matter density in the FRW background is
\begin{equation} \label{EDM}
\rho_D (t)=\dfrac{\alpha_0 \mu_0}{8\pi} \int^{t}_{0} \dd{t'} \int^{l(t')}_{0} \rho_B(t')   \dfrac{1+\mu_0 r}{r^2} e^{-\mu_0 r} a^3(t') 4\pi r^2 \dd{r} \ .
\end{equation}
The $l(t)$ is the length of the spatial effective memory, where the present gravity is not affected by the regions out of it. In other words, it is spatial horizon.
There could be various choices for $l(t)$ such as cosmological horizon $d_p(t) \equiv \int^{t}_{0} a^{-1}(\tau) \dd{\tau}$ or the co-moving Hubble radius $d_c(t) \equiv (a(t)H(t))^{-1}$. For $l(t)=d_p(t)$ the \eqref{EDM} reduces to
\begin{equation} \label{EDM-1}
\rho_D (t) =\dfrac{\alpha_0}{2} \int^{t}_{0}   \rho_B(t')  a^3(t') \left(2- e^{-\mu_0 d_p(t')}  (2+\mu_0 d_p(t'))\right) \dd{t'} \ .
\end{equation}
This element should follow the dynamic of the cosmic fluid. Therefore, the $\rho_B(t)$ and $a(t)$ are the solutions of the
\begin{align}
&\dot{\rho}_B(t) + 3 H(t) \rho_B(t)(1 + w_B) = -\dot{\rho}_D(t) - 3 H(t) (1 + w_D)\rho_D(t) \label{NL Ceq} \\
&\left(\dfrac{\dot{a}(t)}{a(t)}\right)^2 -\dfrac{8\pi G}{3} \rho_B(t) = \dfrac{8\pi G}{3} \rho_D(t) \ . \label{NL 1Feq}
\end{align}
The Hubble parameter $H(t)$ in the first equation, can be replaced from the second one. For $\alpha_0 \ll 1$ the \eqref{NL 1Feq} reduces to
\begin{equation}
\dfrac{\dot{a}(t)}{a(t)} \simeq \sqrt{\dfrac{8\pi G}{3}} \rho_B(t)^{1/2} \left(1 + \frac{1}{2} \frac{\rho_D(t)}{\rho_B(t)} \right) \ . \label{main s eq}
\end{equation}
Therefore,
\begin{multline}
\dot{\rho}_B(t) + 3\sqrt{\dfrac{8\pi G}{3}} \rho_B(t)^{3/2} \left(1 + \frac{1}{2} \frac{\rho_D(t)}{\rho_B(t)} \right) (1 + w_B) = \\
 -\dot{\rho}_D(t) - 3 \sqrt{\dfrac{8\pi G}{3}} \rho_B(t)^{1/2} \rho_D(t) \left(1 + \frac{1}{2} \frac{\rho_D(t)}{\rho_B(t)} \right) (1 + w_D) \ .
\end{multline}
The assumption $\alpha_0 \ll 1$ is consistent with the Newtonian limit which the model based on it. The $\rho_D(t)$ is an integral function of $(a(t),\rho_B(t))$. Since, there is no straight method to solve this integro-differential equation,  some approximation methods would be useful. To this end, first assume that the nonlocal effects are source of perturbation. Therefore, we rewrite the \eqref{NL Ceq} and \eqref{NL 1Feq}, such that the terms up to first order of the nonlocal parameter $\alpha \ll 1$ lives at the RHS as a source,
\begin{equation}
\tilde{\rho}'_B(T) + 3\sqrt{\dfrac{8\pi}{3}}  K_0 \tilde{\rho}_B(T)^{3/2} (1 + w_B) =\mathcal{F}^{I} (T) \label{main c eq}
\end{equation}
where $ K_0=  \sqrt{G \rho_0} t_0$ and
\begin{equation}
\mathcal{F}^{I} (T) =-\tilde{\rho}'_D(T) - 3 \sqrt{\dfrac{8\pi}{3}}  K_0 \sqrt{\tilde{\rho}_B(T)} \tilde{\rho}_D(T) (1 + w_D) - \frac{3}{2} \sqrt{\dfrac{8\pi }{3}} K_0 \sqrt{\tilde{\rho}_B(T)} \tilde{\rho}_D(T) (1+w_B) \ .
\end{equation}
Simultaneously the dimensionless parameters
\begin{align}
& T = \frac{t}{t_0}  \quad\quad\quad \text{where} \quad t_0 = (c\mu_0)^{-1 } \nonumber \\
& \tilde{\rho}_B (t) =\frac{\rho_B(t)}{\rho_0}  \nonumber \\
& \tilde{H} = \frac{a'(t)}{a(t)}  \quad\quad\quad \text{where} \quad a' = \dv{a}{t} \dv{t}{T}   
\end{align}
are defined. 

Up to the zeroth order of iteration, the RHS of the \eqref{main c eq} vanishes and the $(\rho_{B}^{(0)}(T), a^{(0)}(T))$ be same as their classical local counterparts,
\begin{align}
& \rho_{B}^{(0)} = \frac{1}{6 \pi K_0^2 T^2}   \\
& a^{(0)}(T) =  T^{2/3}
\end{align}
where the \eqref{main s eq} is used for derivation of the scale factor. In addition, we consider the condition that the zeroth order solutions treat as well as the classical local solution in order to fix the integration constant.

To find the next order, the $\mathcal{F}^{I} (T)$ is derived from zeroth orders $(\rho_{B}^{(0)}(T), a^{(0)}(T))$. In our interested deep matter-dominated era 
\begin{equation}
\mathcal{F}^{I} (T) =-\tilde{\rho}'_D(T) - \frac{9}{2} \sqrt{\dfrac{8\pi }{3}} K_0 \ \tilde{\rho}_B(T)^{1/2} \tilde{\rho}_D(T) 
\end{equation}
where $(w_B, w_D)=(0,0)$ and the prime denotes the derivative with respect to the $T$.

\begin{figure}
    \centering
		\includegraphics[width=0.6\textwidth]{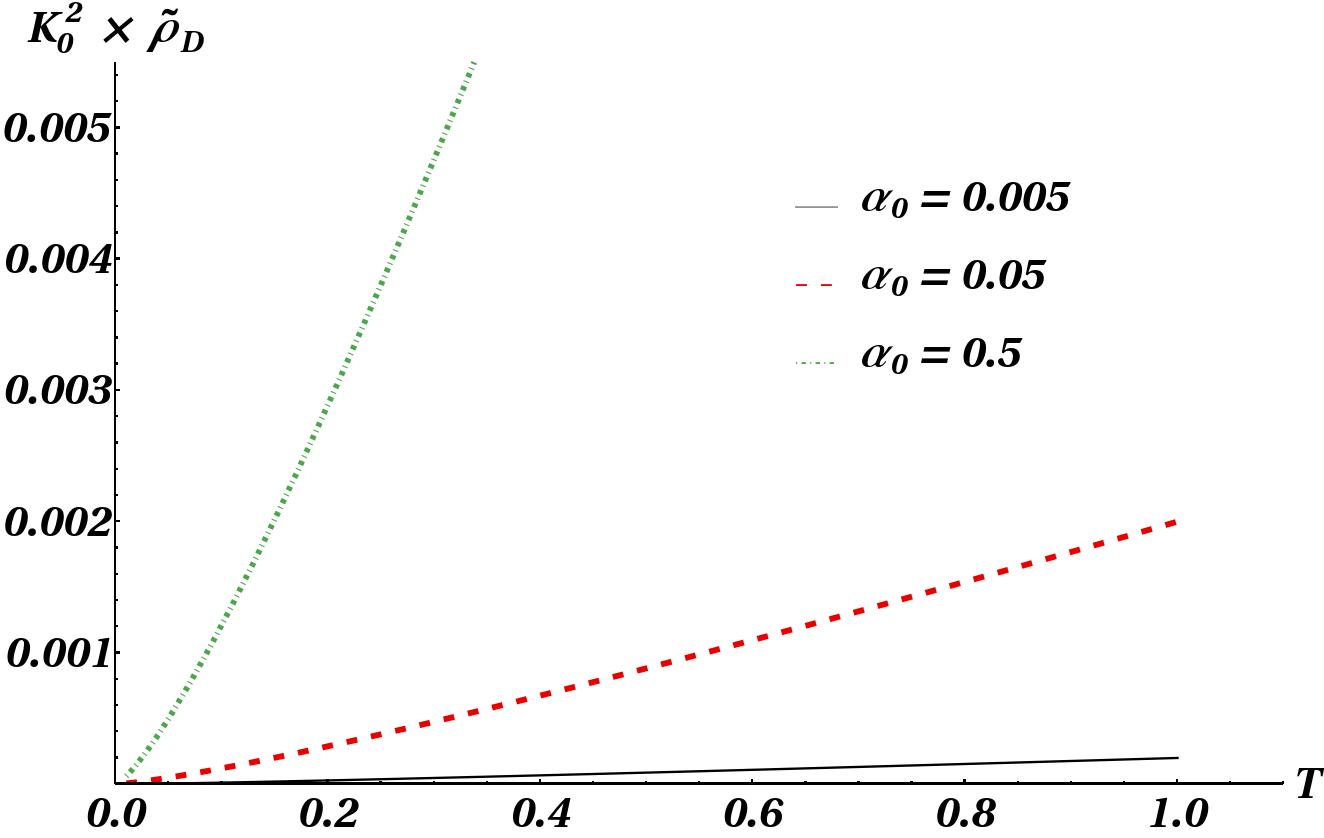}
	\caption{\small{The magnitude of coefficient  $\alpha_0$ would determine the increase in time of the nonlocal effects.}}
	\label{Fig2}
\end{figure}

Up to the first order of the iteration
\begin{equation}
\tilde{\rho}_D \equiv \tilde{\rho}_D^I (T) = \dfrac{\alpha_0}{2} \int^{T}_{T_i}   \rho_{B}^{(0)}(\tau)  a^{(0) \ 3}(\tau) \left(2- e^{-\mu_0 d_{p}^{(0)}(\tau)}  (2+\mu_0 d_{p}^{(0)}(\tau)) \right) \dd{\tau} 
\end{equation}
and
\begin{equation}
d_{p}^{(0)}(T) = \int^{T}_{T_i} \dfrac{\dd{\tau}}{a^{(0)}(\tau)} \ .
\end{equation}
As a result
\begin{align}
& \tilde{\rho}_D^I (T) = \frac{1}{12\pi K_0^2} \left(-\frac{10}{9} + 2T + e^{-3 T^{1/3}} (\frac{10}{9} + \frac{10}{3} T^{1/3} +5 T^{2/3} + 3T)  \right) \alpha_0   \ , \\
& \mathcal{F}^{I} (T) = -\frac{1}{36\pi T K_0^2} e^{-3 T^{1/3}} \left(10 + 30T^{1/3} + 45 T^{2/3} + 21T - 9 T^{4/3} + 2 e^{3 T^{1/3} (-5+12T)} \right) \alpha_0 \ . \label{iteration source}
\end{align}
Based on figure \ref{Fig2}, it is clear that a smaller  $\alpha_0$ value corresponds to a smaller effective density of dark matter.
Moreover, the observed increase in $\tilde{\rho}_D (T)$ over time contradicts the predictions of the standard model of cosmology. This can be explained by considering the nonlocal effects as a history factor. It is anticipated that the size of the causal region connected to a particular event would expand with time, resulting in a corresponding increase in the effective dark matter. Consequently, the effective dark matter as a history would also increase.

To find the corrections on the scale factor, substituting the \eqref{iteration source} in \eqref{main c eq} leads to a complicate nonlinear first order equation which cannot be solved analytically.  We simplified the problem in two steps.

\textit{First}, rewriting the  \eqref{main c eq} for $\mathcal{Z}(T) \equiv \sqrt{\tilde{\rho}_B} (T)$ would lead to
\begin{equation}
\mathcal{Z}'(T) + \frac{3}{2}\sqrt{\dfrac{8\pi}{3}}  K_0 \mathcal{Z}(T)^{2}  =\frac{1}{2}\frac{\mathcal{F}^{I} (T)}{\mathcal{Z}(T)} \ . 
\end{equation}
The zeroth order of $\mathcal{Z}(T)$ is
\begin{equation}
\mathcal{Z}^{(0)}(T) = \frac{1}{\sqrt{6\pi} \ K_0 \ T} \ .
\end{equation}

\begin{figure}
    \centering
		\includegraphics[width=0.6\textwidth]{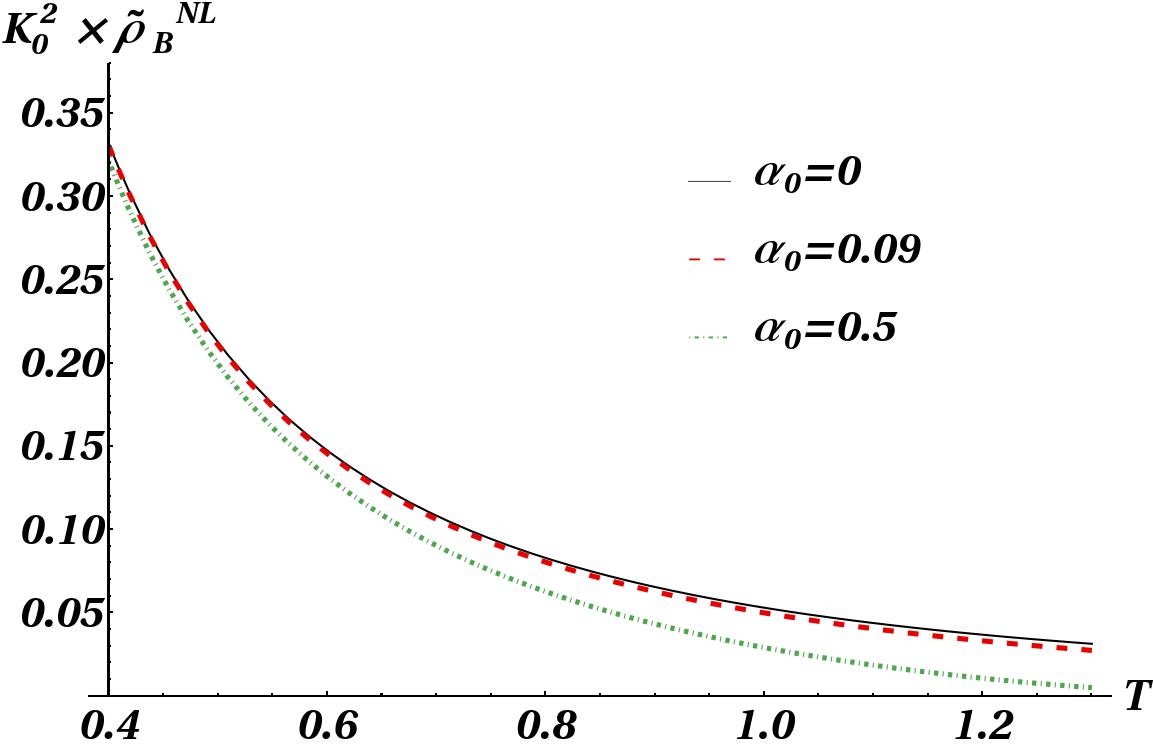}
	\caption{\small{In this model, the Baryonic matter decreases in time in the matter-dominated era.}}
	\label{Fig3}
\end{figure}

\textit{Second}, simplifying $\mathcal{F}^I(T) /2 \mathcal{Z}^{(0)}(T)$ more. To this end the dominant term in \eqref{iteration source} which determines the main behavior of the $\mathcal{F}_I(T)$ is chosen. It is straightforward to see that the main behavior of the iteration source $\mathcal{F}^I(T) /2 \mathcal{Z}^{(0)}(T)$ at the end of the interval $T:(0,1)$ is determined by
\begin{equation}
\frac{1}{2}\frac{\mathcal{F}^{I} (T)}{\mathcal{Z}(T)}  \sim -\sqrt{\frac{2}{3\pi}} \frac{T}{K_0}\alpha_0 \ .
\end{equation}
Thus, the solution of the
\begin{equation}
\mathcal{Z}^{(1)'}(T) + \frac{3}{2}\sqrt{\dfrac{8\pi}{3}}  K_0 \mathcal{Z}^{(1)}(T)^{2}  =-\sqrt{\frac{2}{3\pi}} \frac{T}{K_0}\alpha_0
\end{equation}
is
\begin{equation}
\mathcal{Z}^{(1)}(T) = \frac{J_{-1/3} \left( \mathcal{T}_T \right) c_1  + \sqrt{2} T^{3/2} \Bigl( -2J_{-2/3} \left(\mathcal{T}_T  \right) -J_{-4/3} \left(\mathcal{T}_T \right)c_1 +J_{2/3} \left(\mathcal{T}_T  \right)c_1   \Bigr) \sqrt{\alpha_0} } {2 \sqrt{6\pi} \ K_0 \ T   \Bigl( J_{1/3} \left( \mathcal{T}_T \right) +J_{1/3} \left( \mathcal{T}_T \right) c_1   \Bigr) }
\end{equation}
where  $\mathcal{T}_T \equiv -\frac{2}{3} T^{3/2} \sqrt{2\alpha_0}$. The choice $c_1 = 0$ would satisfy the classical local solution at $\alpha_0 =0$.
This would end to
\begin{equation}
\mathcal{Z}^{(1)}(T) = - \sqrt{\frac{T \alpha_0}{3\pi}} \ \frac{  J_{-2/3} \left(\mathcal{T}_T   \right) }{ \ J_{1/3} \left( \mathcal{T}_T  \right)} \ . \label{Z}
\end{equation}
The decrease of $\tilde{\rho}_B = \mathcal{Z}^{(1)}(T)^2 $ in time can be followed in the figure \ref{Fig3}.
Clearly,
\begin{equation}
\lim_{\alpha_0 \to 0} \mathcal{Z}^{(1)}(T) = \sqrt{\frac{3}{2\pi}} \frac{\Gamma(\frac{4}{3})}{\Gamma(\frac{1}{3})} \frac{1}{K_0 \ T}
\end{equation}
which is consistent with the local solution.

\begin{figure}
    \centering
		\includegraphics[width=0.8\textwidth]{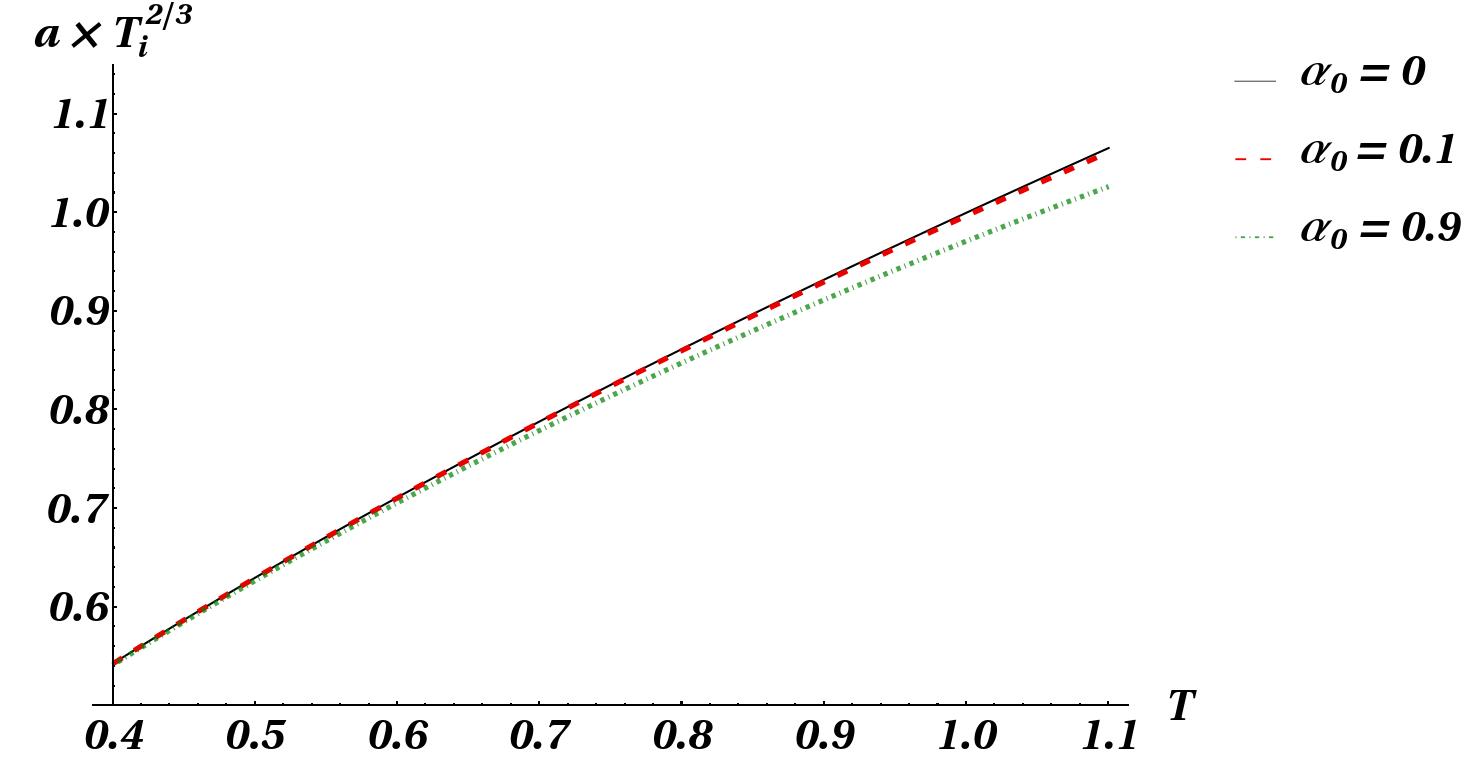}
	\caption{\small{The scale factor evolution in the deep matter-dominated era, would be affected by the nonlocality effects.}}
	\label{Fig4}
\end{figure}

To find the scale factor we use \eqref{Z} to replace 
\begin{equation}
\tilde{\rho}_B(T) \equiv \mathcal{Z}(T)^2 =  \frac{1}{3\pi} \ \frac{  J_{-2/3}^2 \left(\mathcal{T}_T   \right) }{ \ J_{1/3}^2 \left( \mathcal{T}_T  \right)} T \alpha_0 \ .
\end{equation} 
in \eqref{main s eq}. And, as a result
\begin{multline}
a(T) = \left( \frac{T}{T_i} \right)^{2/3} 
 \left( 1- \frac{e^{-3 \ T^{1/3}}}{4374} \left( 24640 + 73920 \ T^{1/3} + 110880 \ T^{2/3} + 110880 \ T + 83160 \ T^{4/3}  
\right. \right. \\
\left. \left. 
 + 49896 \ T^{5/3} + 24543 \ T^{2} + 9477 \ T^{7/3} +2187 \ T^{8/3} + 5 \ e^{3 T^{1/3}} (-4928+81 \ T^2)  \right)\alpha_0 \right) \label{NLscale factor}
\end{multline}
It can be followed in figure \ref{Fig4} that the evolution of the scale factor in this model deviate from the standard one. The enhancement of the nonlocality parameter $\alpha_0$ decreases the expansion rate.

\begin{figure}
    \centering
		\includegraphics[width=0.7\textwidth]{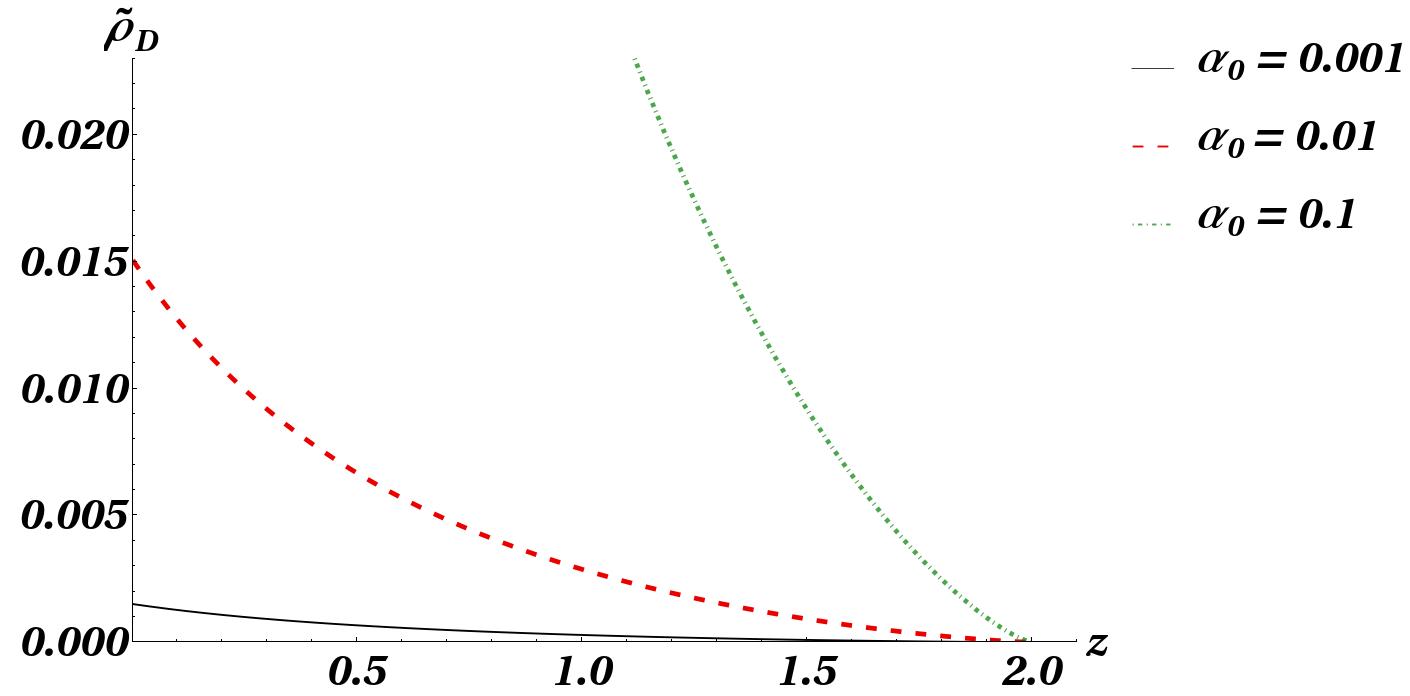}
	\caption{\small{The decrease of the $\tilde{\rho}_D(z)$ as a function of redshift $z$ for any value of $\alpha_0$, clearly is in contrast with $\Lambda$CDM.}}
	\label{Fig5}
\end{figure}

We are interested in $\tilde{\rho}_D(z)$ where $z=\frac{a_0}{a(T)} -1$ is the redshift. The \eqref{NLscale factor} cannot be solved for $T$ simply. It can be shown that the main treatment of the $a(T)$ about $T=1$ is determined by
\begin{equation}
a(T) \sim \left( \frac{T}{T_i} \right)^{2/3} 
 \left( 1- \frac{e^{-3 \ T^{1/3}}}{4374} \left( 2187 \ T^{8/3} \right)\alpha_0 \right)
\end{equation}
where its expansion about the $T=1$ up to the first order is
\begin{equation}
a(T) \sim \frac{2}{9} + \left(\frac{47255}{39366}-\frac{1963921}{78732\ e^3} \right)\alpha_0 + T \left(\frac{8}{9} + \frac{101800}{19683} - \frac{3992723}{39366 \ e^3} \alpha_0  \right)
\end{equation}
and as a result
\begin{equation}
T \sim \left(\frac{2}{9} + \left(\frac{47255}{39366}-\frac{1963921}{78732\ e^3} \right)\alpha_0 -a(T) \right)  \left(\frac{8}{9} + \frac{101800}{19683} - \frac{3992723}{39366 \ e^3} \alpha_0  \right)^{-1} \ .
\end{equation}
Therefore the functional form of $\tilde{\rho}_D(z)$ can be derived. Up to the first order of $\alpha_0$
\begin{equation}
T \sim -\frac{1}{2}+\dfrac{3}{2} a(T) +  \frac{\alpha_0}{2916} \left(13608e^{-3}-1215 +(475975e^{-3} -23020 )a(T) \right)  \ .
\end{equation}
Thus, about $T=1$ we would have
\begin{multline}
\tilde{\rho}_D(z)= \\ \frac{\alpha_0 }{12\pi K_0^2} \left(-\frac{19}{9} + \frac{3}{1+z} +
\frac{e^{-3(\frac{2-z}{2+2z})^{1/3}}}{18} (-7+ \frac{81}{1+z} + 30 \
\sqrt[3]{4 (\frac{2-z}{1+z})} +45 \ \sqrt[3]{2 (\frac{2-z}{1+z})^{2}} ) \right)
\end{multline}
which increases by decreasing the redshift, figure \ref{Fig5}.

\section{Concluding Remarks} \label{fourth}
In this paper, we have studied a nonlocal Newtonian cosmology model in the context of the Mashhoon's nonlocal gravity. This is done in two steps. First, we consider these effects as a source of effective dark matter within the Poisson equation. As a result, the modified Friedman equations \eqref{MF1} and \eqref{MF2} are the ones describe the cosmic fluid evolution. The dynamical system of this model in the effective density parameter space \eqref{parameter space}  closely resembles the $\Lambda$CDM dynamical system. There is only one stable fixed point at $\textbf{P}_{\Lambda}$.

Next, we go beyond the Newtonian cosmology by taking the constancy of the interaction velocity more seriously. 
The assumption $\rho_D(x)=\alpha(t)\rho(x)$ becomes invalid due to the formation of horizons since the nonlocal effects originate from the entire space-time history and not solely the spatial sector. To resolve this contradiction, we redefine the $\rho_D$ by integrating over the entire space-time, as indicated in equation \eqref{our-suggestion}. This results in an increasing density of dark matter with time, which differs from the predictions of the standard model of cosmology. Consequently, the Tohlin-Kuhn kernel is deemed unsuitable if we aim to consider the nonlocal effects as an effective form of dark matter in $\Lambda$CDM.

\vglue2cm
\textbf{Acknowledgements} The author would like to thank Mahmood Roshan and Ali Shojai for useful discussions. This work is supported by a grant from Ferdowsi University of Mashhad.



\begin{thebibliography}{}
\bibitem{Blagojevic}{} M. Blagojevic, \textit{Gravitation and Gauge Symmetries}, London: The institute of Physics (IOP) (2002).
\bibitem{Rahvar}{} S. Rahvar and B. Mashhoon, \href{https://doi.org/10.1103/PhysRevD.89.104011} {Phys. Rev. D \textbf{89}, 104011} (2014).
\bibitem{Mashhoon1990a}{} B. Mashhoon, \href{https://doi.org/10.1016/0375-9601(90)90734-6}{Phys. Lett. A \textbf{143}, 176} (1990).
\bibitem{Mashhoon1993}{} B. Mashhoon, \href{https://journals.aps.org/pra/abstract/10.1103/PhysRevA.47.4498}{Phys. Rev. A \textbf{47}, 4489} (1993).
\bibitem{Bohr}{} N. Bohr and L. Rosenfeld, \href{https://doi.org/10.1103/PhysRev.78.794}{Phys. Rev. \textbf{78}, 794} (1950).
\bibitem{Rovelli}{} C. Rovelli and F. Vidotto, \textit{Covariant Loop Quantum Gravity}, Cambridge: Cambridge University Press (2015).
\bibitem{Mashhoon2011}{} B. Mashhoon, in Cosmology and Gravitation, edited by
M. Novello and S.E. Perez Begliaffa, Cambridge: Cambridge University Press (2011).
\bibitem{Mashhoon Book}{} B. Mashhoon, \textit{Nonlocal Gravity}, Oxford: Oxford University Press (2017).
\bibitem{Chicone2012}{} C. Chicone and B. Mashhoon, \href{https://doi.org/10.1063/1.3702449}{J. Math. Phys. \textbf{53}, 042501} (2012).
\bibitem{Tohline1984}{} J.E. Tohline, \href{https://doi.org/10.1111/j.1749-6632.1984.tb23408.x}{NYASA \textbf{422}, 390} (1984).
\bibitem{Kuhn1987}{} J.R. Kuhn and  L. Kruglyak, \href{https://doi.org/10.1086/164942}{Astro. Phys. J. \textbf{313}, 1} (1987).
\bibitem{Bekenstein1988}{} J.D. Bekenstein, in Proc. of the II Canadian Conf. on General Relativity and Relativistic Astrophysics, edited by A. Coley, C. Dyer and T. Tupper, Singapore: World Scientific (1988).
\bibitem{Mashhoon2022}{} B. Mashhoon, \href{https://doi.org/10.3390/sym14102116}{    Symmetry \textbf{14}, 2116} (2022)
\bibitem{Chicone2016}{} C. Chicone and B. Mashhoon, \href{https://doi.org/10.1063/1.4958902}{J. Math. Phys. \textbf{57}, 072501}  (2016).
\bibitem{Poisson}{} E. Poisson and M.W. Clifford, \textit{Gravity: Newtonian, Post--Newtonian, Relativistic}, Cambridge: Cambridge University Press (2014).
\bibitem{Motahareh}{} M. Kiamari, M. Rahbardar, M. Shokri, and N. Sadooghi, \href{https://doi.org/10.1103/PhysRevD.81.083535}{Phys. Rev. D \textbf{104}, 076023} (2021).
\bibitem{Schutz}{} B. Schutz, \textit{A First Course in General Relativity} (2nd Ed.), Cambridge: Cambridge University Press (2009).
\bibitem{splitting}{} A.H. Barbar, A.M. Awad, and M.T. AlFiky, \href{https://doi.org/10.1103/PhysRevD.101.044058}{Phys. Rev. D \textbf{101}, 044058} (2020).


\end{thebibliography}
\end{document}